\documentclass[letterpaper,twocolumn,10pt]{article}
\usepackage{usenix,epsfig,endnotes}
\begin{document}

\date{}

\title{\Large \bf Tracking Elections: our experience during the presidential elections in Ecuador}

\author{
{\rm Daniel Riofrio}\\
University of New Mexico\\
\and
{\rm Anacaren Ruiz}\\
University of New Mexico
\and
{\rm Erin Sosebee}\\
University of New Mexico
\and
{\rm Qasim Raza}\\
University of New Mexico
\and
{\rm Adnan Bashir}\\
University of New Mexico
\and
{\rm Jed Crandall}\\
University of New Mexico
} 

\maketitle

\thispagestyle{empty}

\subsection*{Abstract}

The world's digital transformation has influenced not only the way we do business, but also the way we perform daily activities. Social media, for instance, has clearly showed us the influence it has on people, especially during social events such as elections. In fact, the past Presidential elections in the United States as well as those in Great Britain (Brexit) and in Colombia (peace agreement referendum) has stablished that social media play an important part in modern politics. In fact, the digital political field expresses in the digital world through political movements and political candidates looking for popular support (\emph{number of followers}), regular citizens' messages discussing social issues (\emph{trending topics flooding social media}), or even political propaganda in favor or against politicians or political movements (\emph{advertisement}).\\

One of the issues with social media in the digital era is the presence of automatic accounts (bots) that artificially fill accounts with fake followers, create false trending topics, and share fake news or simply flood the net with propaganda. All this artificial information may influence people and sometimes may even censor people's real opinions undermining their freedom of speech.\\

In this paper, we propose a methodology to track elections and a set of tools used to collect and analyze election data. In particular, this paper discusses our experiences during the Presidential Elections in Ecuador that were held in 2017. In fact, we show how all presidential candidates prepared an online campaign in social media (\emph{Twitter}) and how the political campaign completely altered the normal subscription of followers. Furthermore, we discuss that the high presence of followers during the period between the first and second round of elections may be altered by automatic accounts. Finally, we use bot detection systems and gathered more than 30,000 bots which were filtered by political motivated content. In our data analysis, we show that these bots were mainly used for propaganda purposes in favor or against a particular candidate.

\section{Introduction}
According to Freedom House, Ecuador's Internet freedom ranks as partially free \cite{FreedomHouse}. Despite the infrastructure investment that the government has made in the past decade, the major problem the country is facing is the control and blockage of content due to copyright infringement targeted to political activists \cite{FreedomHouse}.\\

For the first time in a decade, Ecuador lived a transition in its government. Rafael Correa, former president of Ecuador, for the first time in ten years was no longer a viable candidate for a new presidential period. Therefore, new candidates from several parties participated during the 2017 presidential elections \cite{CNE}.\\

In addition, recent events regarding elections around the world such as the Brexit, the peace agreement referendum in Colombia, and the presidential elections in the United States marked a new political environment in the world due to the influence of social media \cite{SCHMIDBAUER2018148, NBERw23089}. In fact, The New York Times has a series of news regarding how Russia interfered in the 2016 US Presidential Elections through manipulating social media targeting people emotions and preferences \cite{NYTimes}.\\

Hence, the new political environment in the Ecuadorian elections attracted several candidates. Elections in Ecuador were held in two rounds: the first round with 8 candidates on February, 19$^{th}$ 2017, and the second round with two candidates: Lenin Moreno and Guillermo Lasso on April, 2$^{nd}$ 2017.\\

In this work, we show the study we conducted before and during the presidential elections in Ecuador. We present a clear overview of how presidential candidates used the cyberspace to promote their candidacies and the presence of several bots, detected using third party systems for bot detection, such as DeBot \cite{Chavoshi1}, that shared political content against and in favor of candidates.\\ 

Our strategy consisted of three phases: pre-electoral, campaign and post-electoral. The rest of the paper is organized as follows: first we describe how elections in Ecuador work, then we provide a detail explanation of the phases, methodologies and tools used for tracking the elections in the cyberspace (Twitter). We present the results of our Twitter analysis in two scenarios: candidate account analysis and bot detection analysis. Finally, we present a few conclusions and talk about the future work to enhance our approach.


\section{The 2017 Presidential Elections in Ecuador}

Presidential elections in Ecuador are held every four years since the country returned to a democratic system in 1979. Ever since, there has not been any former president reelected, until 2006 where Rafael Correa ran for presidency and stayed in office until 2017. In fact, during 1996 and 2006 the country suffered a complex political instability which caused to have five different presidents in that decade. Hence, tracking the Ecuadorian Elections in 2017 is an important landmark in terms of Ecuador's young democratic system. In addition, these elections are especially important due to the presence of new technologies such as social media and a strong political party, Alianza Pa\'is, which supported Rafael Correa for ten years in office.\\
 
In general terms, presidential elections in Ecuador are mandatory for all Ecuadorian citizens. It consists of two rounds that are held in dates selected by ``Consejo Nacional Electoral'', CNE (the state institution responsible for holding elections in the country). If any candidate is able to obtain more than 40\% of people's popular vote (after correcting for invalid ballots) and if he or she has at least ten percent points over the second is declared President in the first round. Otherwise, the two with higher ballots go for a second round of popular elections, where the one who gets more than 50\% of the ballots is declared President of Ecuador. \\

The 2017 Presidential Elections held in Ecuador took place in February 19$^{th}$, and the second round in April 2$^{nd}$. In the first round, eight parties and political movements inscribed their candidacies: Cynthia Viteri / Mauricio Pozo (Partido Social Cristiano), Abdal\'a Bucaram Pulley / Ramiro Aguilar (Partido Fuerza Ecuador), Iv\'an Espinel / Doris Quiroz (Fuerza Compromiso Social), Guillermo Lasso / Andr\'es P\'aez (Movimiento CREO \& Movimiento SUMA), Len\'in Moreno / Jorge Glas (Movimiento Alianza Pa\'is), Paco Moncayo / Monserratt Bustamante (Izquierda Democr\'atica \& Movimiento Unidad Popular \& Movimiento de Unidad Plurinacional Pachakutik), Washington Pes\'antez / Alex Alc\'ivar (Movimiento Uni\'on Ecuatoriana), and Patricio Zuquilanda / Johnnie Jorgge \'Alava (Partido Sociedad Patri\'otica) \cite{CNE}.\\

After the first round, no candidate obtained the required votes, therefore a second round of elections was held in April 2$^{nd}$ (see table \ref{tbl:firstround}), where Len\'in Moreno was elected President of Ecuador and the runner up was Guillermo Lasso (see table \ref{tbl:secondround}) \cite{CNE}.

\begin{table}[t]
\centering
\begin{tabular}{|l|r|}
\hline
\textbf{Candidate}  & \multicolumn{1}{l|}{\textbf{Votes}} \\ \hline
Len\'in Moreno        & 39.36\%                             \\ \hline
Guillermo Lasso     & 28.09\%                             \\ \hline
Cynthia Viteri      & 16.32\%                             \\ \hline
Paco Moncayo        & 6.71\%                              \\ \hline
Abdal\'a Bucaram      & 4.82\%                              \\ \hline
Iv\'an Espinel        & 3.18\%                              \\ \hline
Patricio Zuquilanda & 0.77\%                              \\ \hline
Washington Pes\'antez & 0.75\%                              \\ \hline
\end{tabular}
\caption{Results of the first round of elections \cite{CNE}.}
\label{tbl:firstround}
\end{table}

\begin{table}[h]
\centering
\begin{tabular}{|l|r|}
\hline
\textbf{Candidate} & \multicolumn{1}{l|}{\textbf{Votes}} \\ \hline
Len\'in Moreno       & 51.16\%                             \\ \hline
Guillermo Lasso    & 48.84\%                             \\ \hline
\end{tabular}
\caption{Results of the second round of elections \cite{CNE}.}
\label{tbl:secondround}
\end{table}


\section{Measuring political content in social media}

From the experience in other elections around the world, and the strong claims by political activists in Ecuador that the government were manipulating public opinion on local social media \cite{EcuadorTransparente, GonzaloBilbao, Fundamedios, SamuelWoolley}, we set our main objective to track political content on social media and measure censorship or interference by automatic accounts (bots) during the presidential elections.\\

We chose Twitter since it is a widely used social media in Ecuador with over 2,000,000 active users \cite{Twitter2015A}. In fact, Twitter has become an important social media that in the past decade, the former President replied to people's requests from his personal Twitter account.\\

In the following subsections, we introduce the methodology used, the tools we used to collect bot information, and the tools we developed to analyze the data gathered from January to April 2017.

\subsection{Methodology}

Our methodology consists of dividing the electoral year in phases: pre-electoral, campaign and post-electoral.
In the case of the 2017 Presidential elections in Ecuador, we starting planning our experiments in August 2016. The main problem we encountered is that measuring anything in social media requires establishing a base line. Therefore, we designed the pre-electoral phase in order to collect all data in advance before any political party or political movement made an official announcement of their presidential candidate. We selected over 100 special accounts and started following their activity on Twitter.\\

\begin{figure*}[h]
\centering
\includegraphics[scale=0.3]{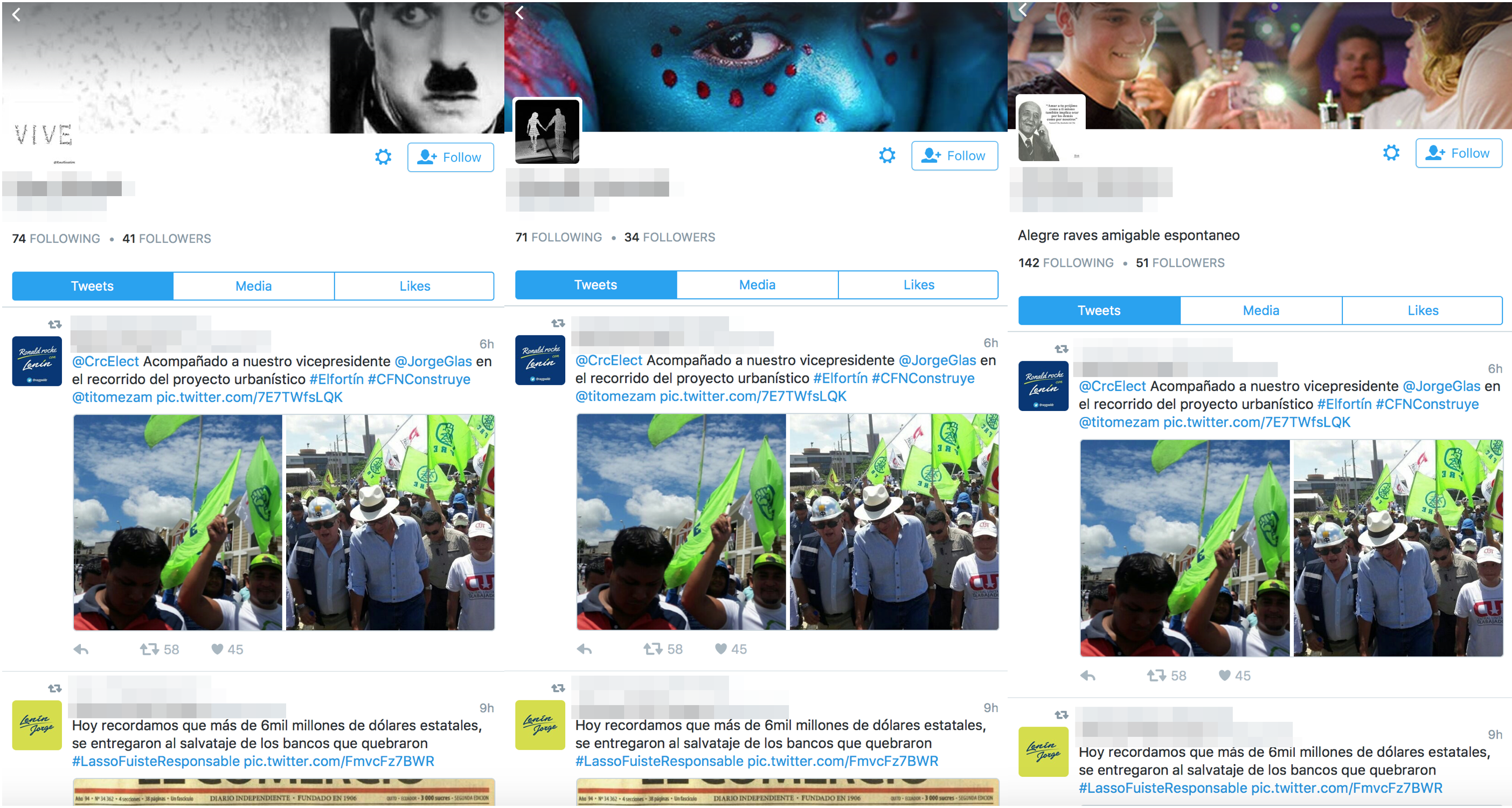}
\caption{Example of automatic accounts detected by DeBot during the 2017 Presidential Elections in Ecuador.}
\label{fig:Bots}
\end{figure*}

We planned two main experiments: data recollection for bot analysis, and data recollection for special accounts analysis. For the first experiment, we collected and classified data from November to December 2016 related to political events in the country. We filtered the trending topics as well as the most common words used in social media to describe political events and generated a list of potential political topics which included potential candidate names, name of political parties and movements, name of political scandals, etc. This list was used into the Twitter API to feed DeBot, which is the tool we used to search for automatic accounts. DeBot is an unsupervised bot classifier developed by Chavoshi et al.\cite{Chavoshi1, Chavoshi2}. DeBot works in two phases: first, it listens to users mentioning a list of keywords and it creates a set of potential bots based on each account behavior. Second, it listens the activity of each potential bot for a few hours and creates a signal based on its activity. Later, DeBot calculates the correlation among accounts and determines which ones are related and place them in clusters \cite{Chavoshi2}. Please see figure \ref{fig:Bots} where we show three political bots promoting a particular candidate during the campaign.\\

Our second experiment consisted of tracking all special accounts on Twitter. We collected snapshots of Twitter followers from a set of selected accounts (picked through a previous analysis of Ecuador's political situation) conformed by political parties and movements, government institutions, and political personalities (potential candidates, political activists, and active politicians). The graph was collected using the Twitter API and required almost two days in order to traverse the list of accounts without exceeding Twitter rate limits. 

\subsection{Phase 1: Pre-electoral phase}

The pre-electoral phase started in November to December 2016 for our first experiment. We gathered hundreds of trending topics and started to collect bots from January 2017. The base line we established was the list of trending topics related to politics that were analyzed and validated by an Ecuadorian citizen knowledgable in politics. \\

For our second experiment, the base line was established by collecting the first graph in January 2017. As we will see later in our results, the base line in this case was not perfect, because new candidates emerged in the very last month before elections.

\subsection{Phase 2: Campaign}

During the campaign, we kept our bot analyzer running every single day gathering potential bots. We refreshed the database containing trending topics related to politics based on emergent topics that were not considered in our base line. \\

For our second experiment, we gathered a second snapshot of the graph of all special accounts. We also added those accounts that we did not considered in our base line. The second snapshot was taken one day before the first round of elections, February 18$^{th}$ 2017. 

\subsection{Phase 3: Post-electoral phase}

For our post electoral phase, we kept DeBot running until the end of April 2017. And, we gathered a third snapshot of all followers of our special accounts, also by the end of April.

\section{Results}

We present our results in two sections: bot analysis and twitter graph analysis. 

\section{Bot Analysis}

\begin{figure}[h]
\centering
\includegraphics[scale=0.23]{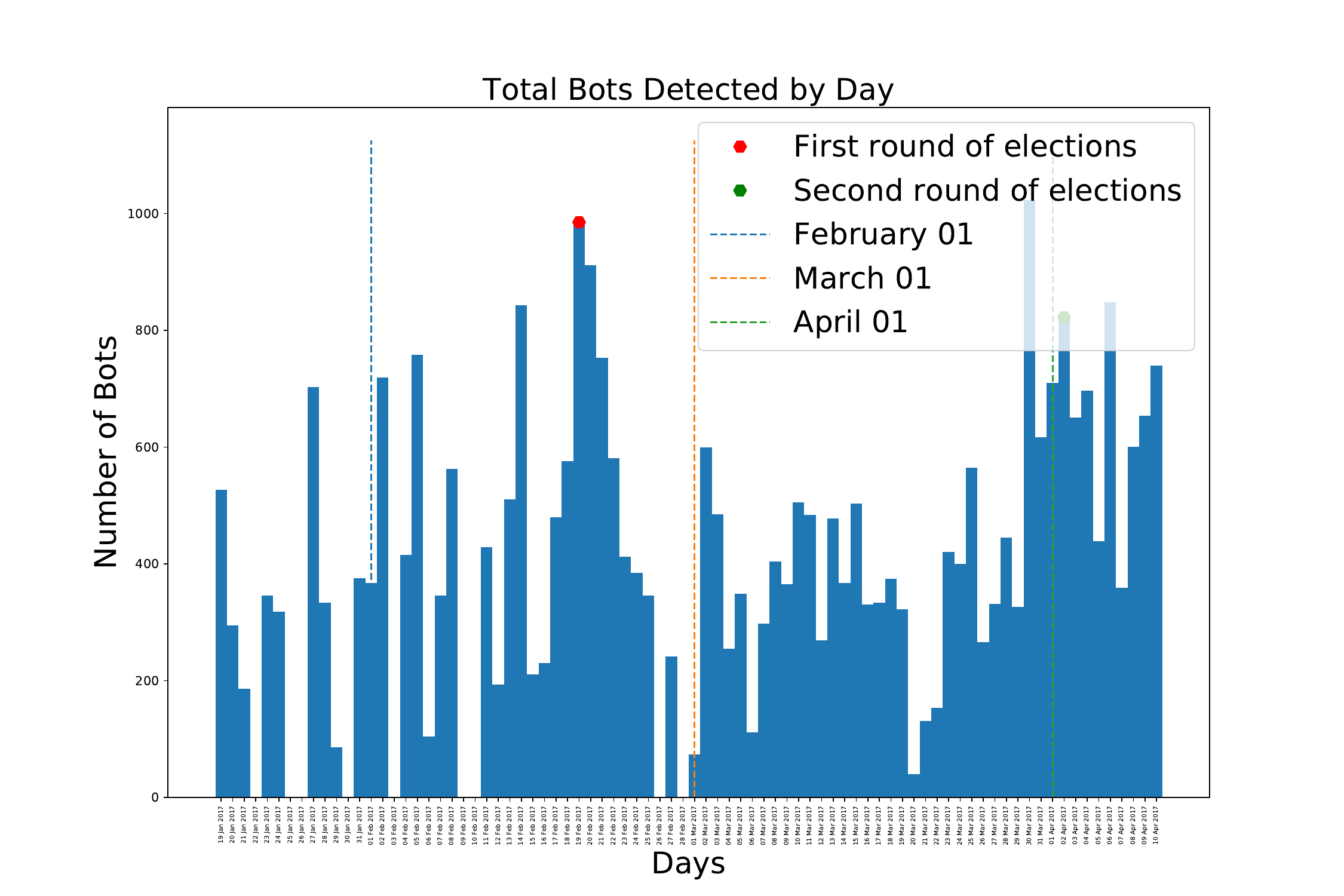}
\caption{Total political bots detected during the 2017 Presidential Elections in Ecuador.}
\label{fig:AA}
\end{figure}

\begin{figure}[h]
\centering
\includegraphics[scale=0.40]{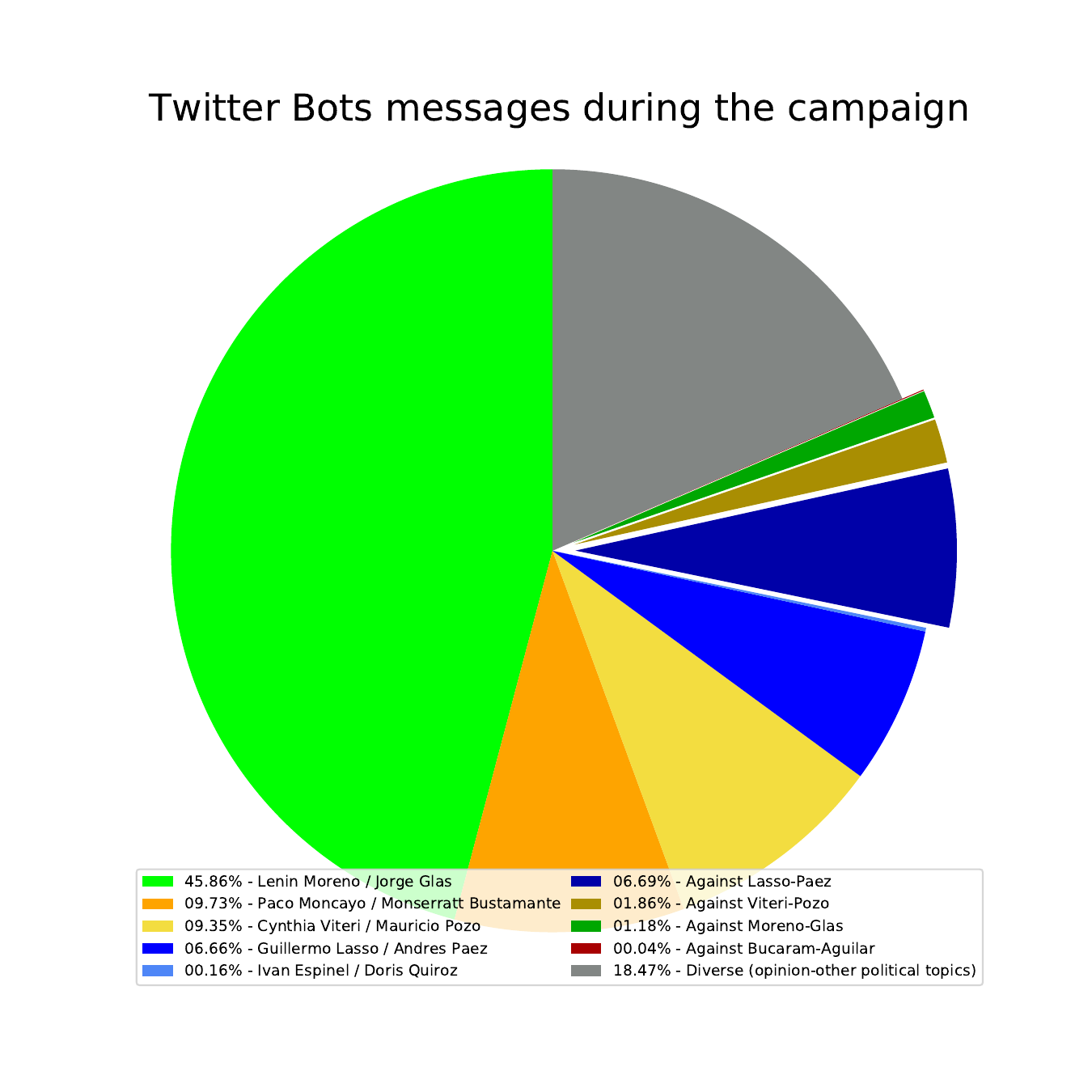}
\caption{Twitter bots messages during the campaign. Lighter colors represent positive messages about a candidate, darker colors represent negative messages about a candidate.}
\label{fig:AB}
\end{figure}

DeBot collected 32,672 bots from January to April 2017. Figure \ref{fig:AA} show the distribution of these bots every day from January to April 2017. These bots interacted more in business hours and collectively increased their activity close to the election days. In particular, we ran a classification analysis on all the collected bots in order to understand what they were promoting during the campaign. We gathered all keywords used by all the tweets we were able to collect from each bot and match positive keywords in favor to a candidate and negative keywords against of a candidate. This classification was performed by ranking the keywords by which were the most mentioned to the least and having our expert validate whether a keyword was in favor or against a certain candidate. We later clustered all tweets produced by bots in these categories and generated figure \ref{fig:AB}.

\section{Graph Analysis}

For this experiment, we considered the differences among graphs in time. We calculated all new accounts following each candidate from the pre-electoral phase into the campaign, and from the campaign into the post-electoral phase. We found the most change in the first graph difference. Table \ref{phase2vsphase1} shows the transition that the Twitter accounts of Presidential candidates suffered in the lapse of a month (January to February). We have included former President, Rafael Correa, to the list because, even though he was not a candidate, he had an extremely active set of followers in Twitter. \\

\begin{table*}[h]
\centering
\begin{tabular}{|l|l|r|r|r|r|}
\hline
\textbf{Candidate's Name} & \textbf{User Account} & \textbf{Phase 1 - Followers} & \textbf{Phase 2 - Followers} & \textbf{Difference} & \textbf{Increase \%} \\ \hline
Cynthia Viteri            & @CynthiaViteri6       & 99,669                                     & 117,634                                & 17,965              & 18.02\%              \\ \hline
Dalo Bucar\'am              & @daloes10             & 324,443                                    & 330,189                                & 5,746               & 1,77\%               \\ \hline
Iv\'an Espinel              & @IvanEspinelM         & N/A                                        & 10,287                                 & N/A                 & N/A                  \\ \hline
Guillermo Lasso           & @LassoGuillermo       & 244,990                                    & 259,444                                & 14,454              & 5.90\%               \\ \hline
Len\'in Moreno              & @Lenin                & 4,462                                      & 126,791                                & 122,329             & 2,741.57\%           \\ \hline
Rafael Correa             & @MashiRafael          & 2,877,737                                  & 3,002,662                              & 124,925             & 4,34\%               \\ \hline
Paco Moncayo              & @PacoMoncayo          & 10,352                                     & 22,988                                 & 12,636              & 122.06\%             \\ \hline
Washington Pes\'antez       & @PesantezOficial      & 1,807                                      & 2,021                                  & 214                 & 11.84\%              \\ \hline
Patricio Zuquilanda       & @ZuquilandaDuque      & N/A                                        & 1,562                                  & N/A                 & N/A                  \\ \hline
\end{tabular}
\label{phase2vsphase1}
\caption{Account follower differences among campaign and pre-electoral phases.}
\end{table*}

\begin{figure}[t]
\centering
\includegraphics[scale=0.23]{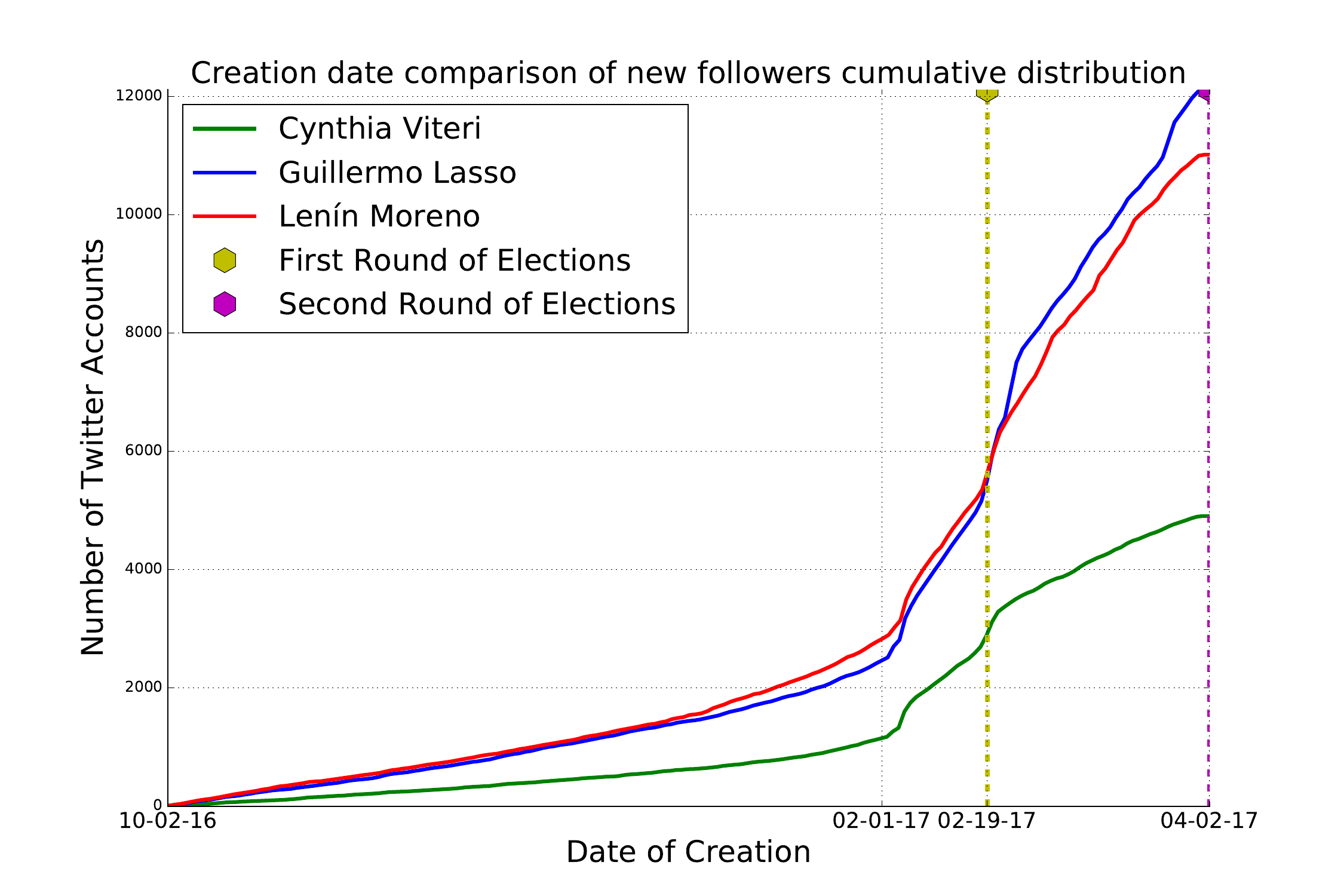}
\caption{Cumulative view: graph difference analysis of new follower accounts' date of creation in time. Blue line shows Guillermo Lasso's twitter account, red shows Len\'in Moreno's twitter account and green shows Cynthia Viteri's twitter account}
\label{fig:BA}
\end{figure}

\begin{figure}[t]
\centering
\includegraphics[scale=0.23]{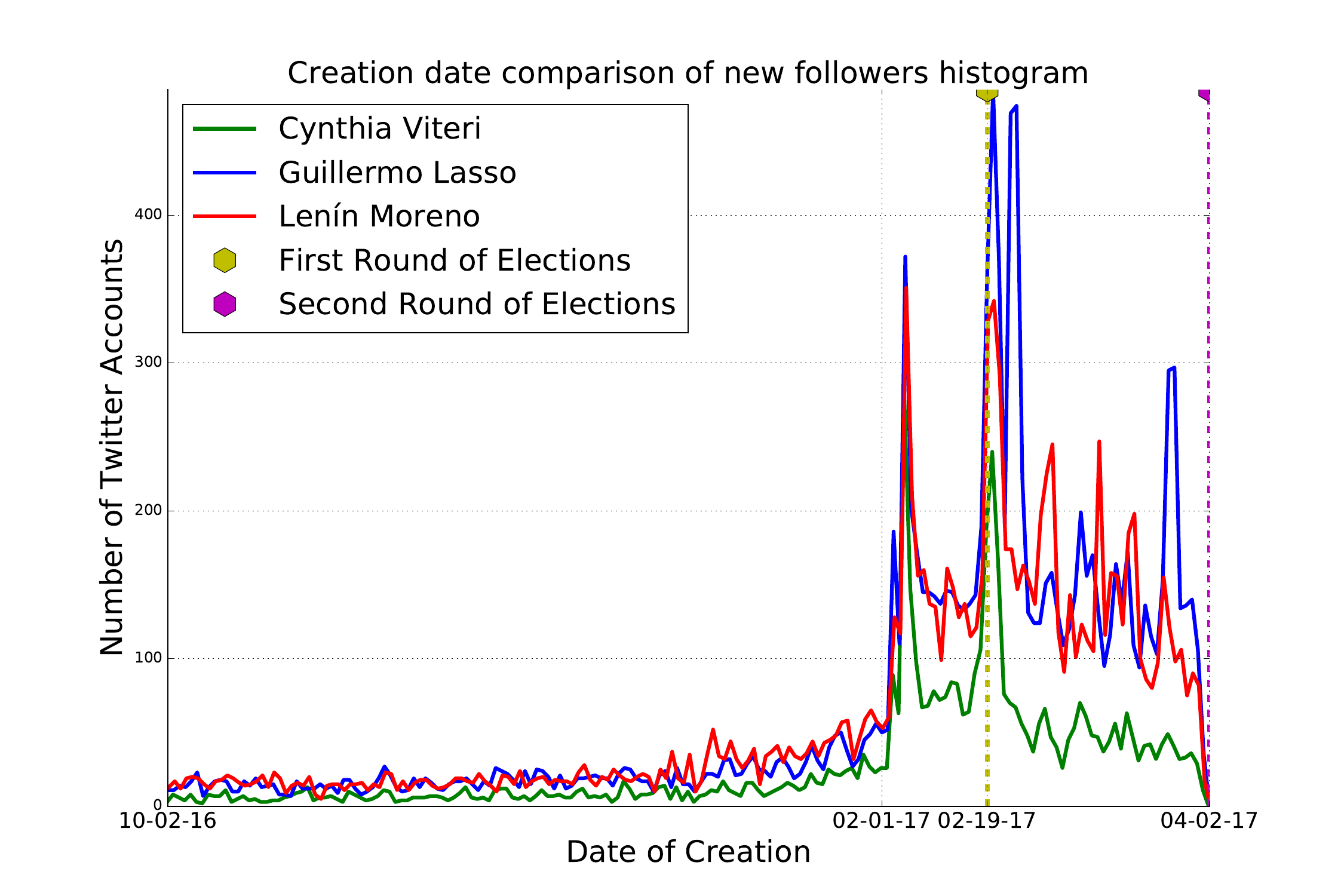}
\caption{Activity view: graph difference analysis of new follower accounts' date of creation in time. Blue line shows Guillermo Lasso's twitter account, red shows Len\'in Moreno's twitter account and green shows Cynthia Viteri's twitter account.}
\label{fig:BB}
\end{figure}

In addition, we compare the three main candidate twitter accounts according to their final vote (see table \ref{tbl:firstround}) and show in both a cumulative distribution of the date of creation of each new follower and observe a particular pattern from the day the official campaign started (February 1$^{st}$ to the second round of elections (April 2$^{nd}$). Figures \ref{fig:BA} and \ref{fig:BB} show how the patterns of the main candidates increase the number of recently created accounts during the campaign. \\

\begin{figure}[t]
\centering
\includegraphics[scale=0.23]{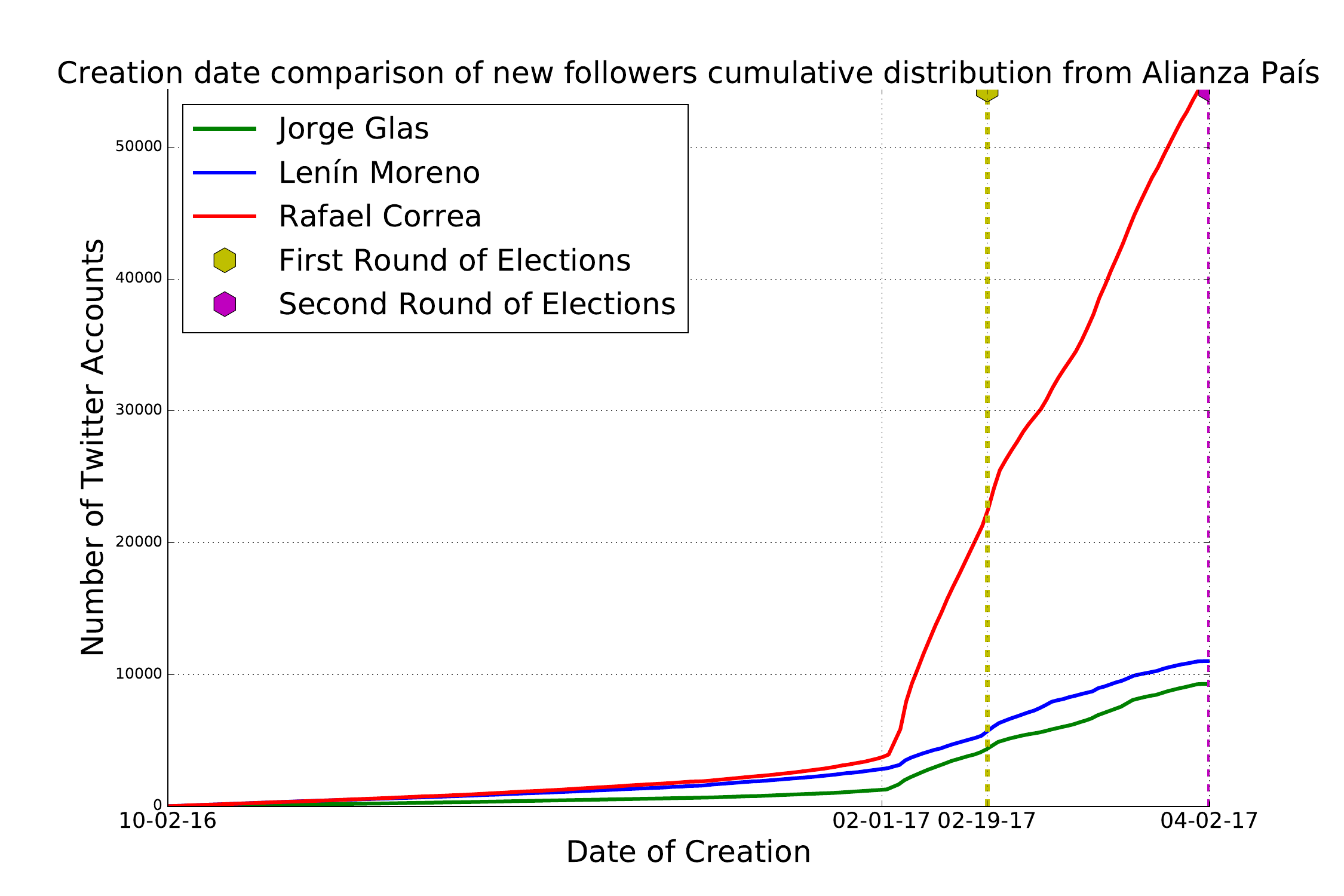}
\caption{Cumulative view: graph difference analysis of new follower accounts' date of creation in time. Red line shows Rafael Correa's twitter account, blue shows Len\'in Moreno's twitter account and green shows Jorge Glas' twitter account.}
\label{fig:BC}
\end{figure}

\begin{figure}[h!]
\centering
\includegraphics[scale=0.23]{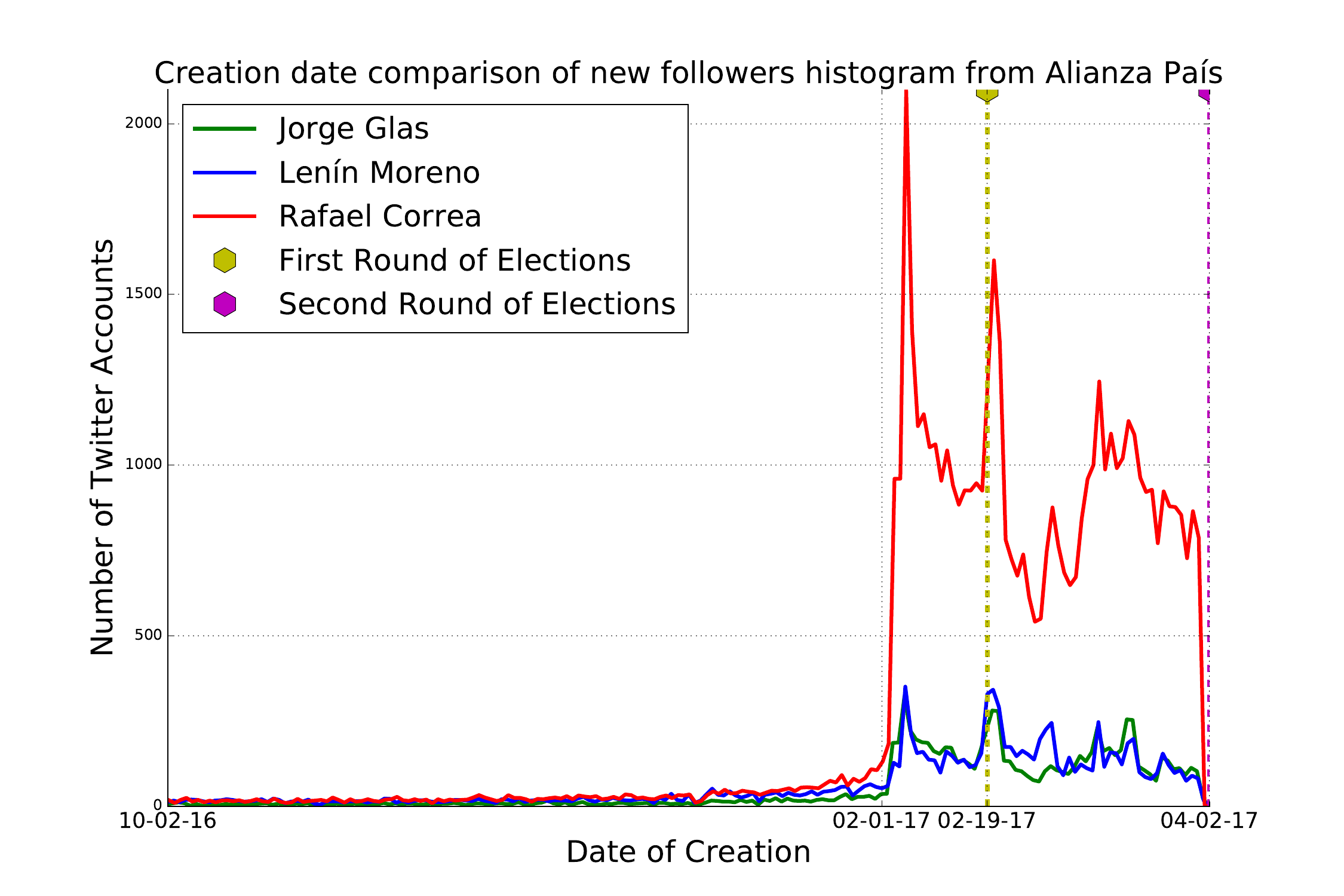}
\caption{Cumulative view: graph difference analysis of new follower accounts' date of creation in time. Red line shows Rafael Correa's twitter account, blue shows Len\'in Moreno's twitter account and green shows Jorge Glas' twitter account.}
\label{fig:BD}
\end{figure}

Finally, we compare the twitter accounts of official candidates (those that belong to the same political movement as former President Rafael Correa) with Rafael Correa's twitter account. We found that even though, Rafael Correa was not a candidate, he had an even higher follower activity during the campaign. In fact, figures \ref{fig:BC} and \ref{fig:BD} show how Rafael Correa twitter account increases the number of recently created accounts during the campaign in about 5 times those from Len\'in Moreno's and Jorge Glas' twitter accounts. 

\section{Conclusions}

For the first time in any election in Ecuador, we present strong evidence that automatic accounts are used by political parties and movements to promote their candidates. We show how these automatic accounts have been used to promote or discredit other candidates. In fact, we show how almost 46\% of all bots collected supported the official candidate, Len\'in Moreno, and that other candidates such as Guillermo Lasso received bad press for almost every single positive tweet in his favor.\\

Additionally, we show with concern how the political campaign flooded Twitter with thousands of accounts that were recently created to support candidates. We understand that once we obtain the difference graph from our campaign data from the pre-electoral phase, our comparisons should accentuate the final period were new activity happened in Twitter, but the increasing slope of new followers in the main candidates and in the account of Rafael Correa alerts about the nature of such accounts and therefore demands further research.\\

Finally, it is clear that all candidates with no exception understood the importance of the digital era in politics. All of them prepared twitter accounts and their accounts increased their followers during the campaign. It is important to mention that even though we gathered data for the post electoral phase, since the campaign and the post-electoral phase were very close to each other, we were unable to see any significant differences.

\section{Acknowledgments}

This work was funded by the Open Technology Fund (OTF) through their Information Controls Fellowship which was held at The University of New Mexico (UNM). Special thanks to Prof. Jed Crandall from UNM and his research group. Last but not least, we would like to thank the civil society groups that supported our work: UsuariosDigitales, Fundacion MilHojas, 4Pelagados, and EcuadorTransparente.\\

{\footnotesize \bibliographystyle{acm}

\bibliography{./dr_bibliography}}


\end{document}